\documentclass{aa}  
\usepackage[switch]{lineno}

\usepackage{graphicx}
\usepackage[dvipsnames]{xcolor}
\bibpunct{(}{)}{;}{a}{}{,}
\usepackage{txfonts}

\usepackage{siunitx}
\usepackage{float}
\begin{document}

   \title{Signatures of localised particle acceleration at a global coronal shock wave}

   \author{C.~Cuddy
          \inst{1, 2},
          D.~M.~Long
          \inst{1, 2}, 
          M.~Nedal
          \inst{2}, 
          S.~Bhunia
          \inst{3},
          P.~T.~Gallagher
          \inst{2}
          }

   \institute{Centre for Astrophysics and Relativity, School of Physical Sciences, Dublin City University, Glasnevin Campus, Dublin, D09 V209, Ireland.
        \and
             Astronomy \& Astrophysics Section, School of Cosmic Physics, Dublin Institute for Advanced Studies, DIAS Dunsink Observatory, Dublin, D15 XR2R, Ireland.
             \and
             LIRA, Observatoire de Paris, PSL Research University, CNRS, Sorbonne Université, Université Paris Cité, 5 place Jules Janssen, 92195 Meudon, France
             }

   \date{}

  \abstract
{Extreme ultraviolet (EUV) waves are global waves in the solar corona which can accelerate particles. The efficiency of this acceleration depends on local plasma characteristics such as the Alfv\'en speed and the geometry of the magnetic field. This shock-driven particle acceleration can produce radio signatures such as Type II radio bursts and herringbone emission.}
{Here we investigate signatures of particle acceleration by a weak coronal shock on 10 March 2024. In particular, we combine EUV images with radio imaging and spectral observations to determine how and where this weak shock could accelerate energetic particles.}
{A potential field source surface extrapolation was used to examine the pre-eruption ambient magnetic field while the evolution of the global wave was probed using running difference and base difference EUV images. The EUV images enabled the speed and Alfv\`en Mach number of the EUV wave to be characterised. The combination of radio images and dynamic spectra provide evidence of beams of shock-accelerated electrons localised to a dimming region at the time the EUV wave passes through it. The speeds and energies of these electrons were estimated from the drift rates of their herringbones.}
{The EUV wave initially propagated due West, channelled by two large loop systems, before changing direction northward. From the EUV intensity jump at the wavefront, the Alfvén Mach number was estimated to be approximately 1.005 at the time that the herringbones were produced. The herringbone drift rates revealed accelerated electron energies of 75.32 to 122.10~keV, using Newkirk density models with scaling factors of 1.3 to 2.6.}
{These observations suggest that the weak lateral shock impacted quasi-perpendicular open field in a dimming region, enabling localised particle acceleration. This indicates that the geometry of the ambient magnetic field relative to the shock strongly governs where particles can be accelerated.}

\titlerunning{Particle acceleration at a coronal shock wave}
\authorrunning{C.~Cuddy et al.}
   \maketitle

\section{Introduction}
\citet{1968SoPh....4...30U} proposed the existence of coronal fast mode magnetohydrodynamic waves as the cause of Moreton-Ramsey waves, chromospheric bright fronts in H-alpha images, first observed by \citet{1960AJ.....65U.494M,1960PASP...72..357M}. Uchida suggested that a Moreton-Ramsey waves were the footprints of coronal waves interacting with the chromosphere, to explain their observed high speeds (500–1500~\si{\kilo\metre\per\second}), which were incompatible with wave propagation through the chromosphere directly but consistent with estimated coronal fast-mode speeds. Globally propagating waves in the solar corona were first observed by the Extreme ultra-violet Imaging Telescope \citep[EIT;][]{1995SoPh..162..291D} onboard the Solar and Heliospheric Observatory (SOHO) in 1997 \citep{1998GeoRL..25.2465T, 1998HiA....11..861D}. They were observed as a bright circular wavefront propagating out from an active region, across the disk, at a height of approximately 70-90~Mm above the photosphere \citep[cf.][]{2009SoPh..259...49P}. It was distinguished by \citet{2002ApJ...569.1009B} that these EUV waves are always accompanied by a CME but not every CME has an associated EUV wave. From initial studies based on EIT observations, the speeds of EUV waves were found to be typically in the range of 200-400~\si{\kilo\metre\per\second} \citep{2009ApJS..183..225T}, much lower than measured Moreton-Ramsey wave speeds and expected coronal fast mode wave speeds.

Due to this apparent discrepancy, other theories arose to explain EUV waves, with some describing these phenomena as slow-mode solitons \citep{2007ApJ...664..556W}, and some describing them as pseudo-waves. Theories in the pseudo-wave branch postulate that these brightenings arise from field line stretching \citep{2002ApJ...572L..99C}, Joule heating in current shells \citep{2007A&A...465..603D}, or continuous small scale reconnection \citep{2007ApJ...656L.101A}. After the launch of Solar Dynamics Observatory \citep[SDO;][]{2012SoPh..275....3P}, its EUV imager, Atmospheric Imaging Assembly \citep[AIA;][]{2012SoPh..275...17L} enabled high cadence imaging of the solar corona, with a time resolution of 12s in comparison with EIT's 720s. Studies of EUV waves imaged with AIA have yielded much higher EUV wave speeds than previous works. \citet{2013ApJ...776...58N} show that EUV wave speeds span a broad range of approximately 200–1500~\si{\kilo\metre\per\second}, with mean and median values around 600~\si{\kilo\metre\per\second}. These speeds, results of differential emission measure analysis \citep{2015ApJ...812..173V} and simulations \citep{2012ApJ...750..134D, 2021ApJ...911..118D}, and the observation of co-spatial radio sources, \citep[cf.][]{2013NatPh...9..811C, 2019NatAs...3..452M} support the interpretation of EUV waves as brightening due to heating and compression at the fronts of fast mode magnetohydrodynamic waves or shocks, driven by the expansion of the associated CME. As a result, there is a growing general consensus in favour of the interpretation of EUV waves as fast mode waves or shocks \citep{Long2017}.

{Radio sources at the EUV wave front reveal locations and times where the wave exceeds the local fast mode speed, becoming a shock and accelerating particles \citep[cf.][]{2019NatAs...3..452M}. Combining this with radio spectra, we can estimate the height where electrons were accelerated by the shock, and their resultant energy. Charged particles can become accelerated by shocks via shock drift acceleration and diffusive shock acceleration, \citep{1983ApJ...267..837H, 1994PASA...11...21S}. Coronal shocks have been shown to accelerate particles \citep{2012ApJ...752...44R, 2014SoPh..289.1731P}, despite typically being subcritical \citep{2021ApJ...921...61L}, suggesting that even weak shocks can accelerate particles in coronal conditions. Scattering and turbulence in the corona can play an important role, especially in the case of diffusive shock acceleration, where the energy gained by a particle in one crossing of the shock is very small but after many repeated crossings the particle can become highly accelerated \citep{2001PASA...18..361B, 2013NatPh...9..811C}. Shocks can become efficient at energising electrons via shock drift acceleration in regions where the magnetic field is quasi-perpendicular to the shock normal, due to the motion of the electrons being confined along the shock \citep{2013NatPh...9..811C, 2020A&A...633A..56M}. \citet{2021ApJ...921...61L} attributed the acceleration of electrons by a weak shock to interaction between the shock and the ambient magnetic field.

Type II and type III radio bursts are signatures of accelerated electrons in the corona \citep{1985srph.book..333N, 2020FrASS...7...56R}. When electrons are accelerated in coronal plasma either by shocks or magnetic reconnection, this non-thermal population of electrons causes electrostatic plasma oscillations called Langmuir waves. These oscillations can combine with ion acoustic waves, producing electromagnetic waves at the fundamental frequency of the plasma. Two Langmuir waves can alternatively coalesce to produce electromagnetic waves at the second harmonic of the plasma frequency. The fundamental plasma frequency is proportional to the square root of the electron number density. At the density range of the corona, the range of fundamental and harmonic plasma frequencies are in the range of tens to hundreds of MHz.

Type III radio bursts are short-lived radio bursts with a large negative drift rate (typically $df/dt \approx$ -100~\si{\mega\hertz\per\second} \citep{reid2014review}). They represent plasma emission at rapidly decreasing frequency due to a beam of energetic electrons escaping on open magnetic field lines. They are regularly observed from active regions during flares, in which case they signify the escape of electrons which have been accelerated by magnetic reconnection \citep{reid2014review, 2020FrASS...7...56R}. Type III bursts have also been observed near the location of a shock front, as inferred from an observed EUV wave in \citet{2021ApJ...921...61L}. In this case the emission indicates the escape of shock accelerated electrons. Type II radio bursts are longer-lived radio bursts with lower drift rates (typically $df/dt \approx$ 0.1~\si{\mega\hertz\per\second} \citep{1996A&AS..119..489M}). They occur due to shock acceleration of electrons as a shock travels outward and passes through plasma of progressively lower electron density. Type II radio bursts have also been imaged at EUV wave fronts \citep[cf.][]{2021A&A...651L..14M}.

Herringbones are a fine structure that can be present in type II bursts \citep{2005A&A...441..319M}. They appear above and/or below the type II on a dynamic spectrum, with those extending to lower frequencies having a forward drift and those extending to higher frequencies having a reverse drift. They are caused by shock accelerated electrons travelling on magnetic field lines, either out into interplanetary space (forward drift) or down deeper into the corona/chromosphere (reverse drift). Radio imaging of herringbone radio emission has confirmed that they originate from shocked regions \citep{2019NatAs...3..452M}.

In this paper, we combine observations of an EUV wave from SDO/AIA with spectral observations a type II radio burst and herringbones from Radiospectrographic Observations for FEDOME and the Study of Solar Eruptions \citep[ORFEES;][]{2021JSWSC..11...57H} and Compound Astronomical Low frequency Low cost Instrument for Spectroscopy and Transportable Observatory) \citep[CALLISTO;][]{benz}, along with radio images from Nançay RadioHeliograph \citep[NRH;][]{1988AdSpR...8k.193M}. We present rare imaging observations of herringbones, which interestingly appear at the EUV wavefront as it interacts with a transient dimming region. From our analysis of the AIA images, we find that the shock responsible for particle acceleration at this location was very weak. Our findings suggest that the interaction between the shock and the local magnetic field was a crucial factor in enabling shock acceleration of electrons in this event. This is in agreement with the findings of \citet{2021ApJ...921...61L}. In Section 2 we detail the observational data utilised in this study, in Section 3 we describe our methods and results, and in Section 4 we provide further discussion and conclusions.

\begin{figure*}[!t]
    \centering
    \includegraphics[width=\textwidth]{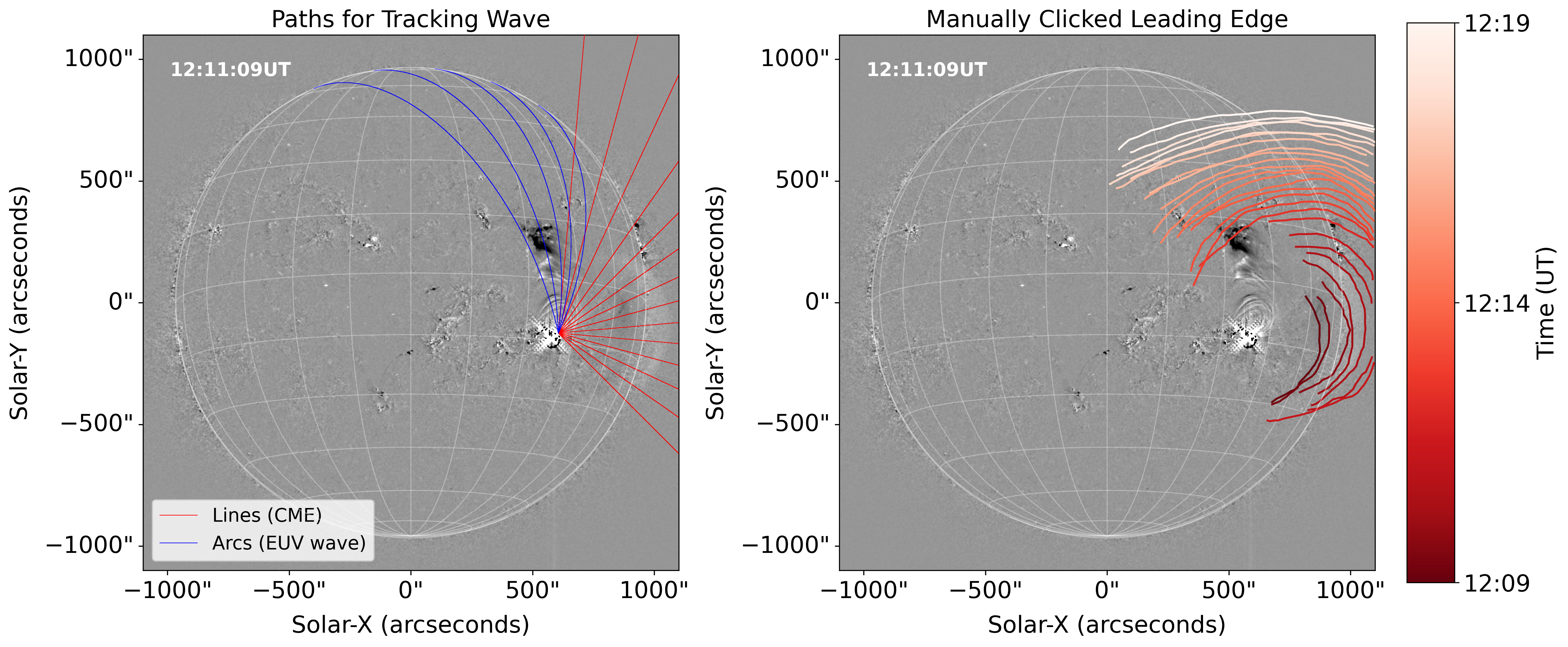}
    \caption{AIA 211~{\AA} running difference images from 12:11:09~UT on 10 March 2024. Left: Paths for tracking the EUV wave plotted on a 1 minute offset AIA 211~{\AA} running difference image. Right: For each of the 27 running difference images made between 12:09:33~UT and 12:19:33~UT, there is a front shown here in a unique shade of red, which connects the manually selected points along the leading edge of the EUV wave. In the online version we present this information in a movie.}
    \label{fig:paths_and_clicks}
\end{figure*}

\section{Observations}

On 10 March 2024 there was a (Geostationary Operational Environmental Satellite) GOES M7.4 class X-ray flare detected from AR13599, between 12:00~UT and 12:20~UT, peaking at 12:08:54UT. The associated CME reached the base of the LASCO C2 field of view (2.5~$R_{\odot}$) at 12:36~UT, with a central position angle of 275\textdegree\ counter-clockwise from North, and a velocity of 307~\si{\kilo\metre\per\second}. Images from AIA show remote brightening in AR13602 at the time of the flare, indicating that this active region is magnetically connected to the flare site. AIA running difference images depict an EUV wave propagating out from AR13599 between 12:09:33~UT and 12:19:33~UT. AIA base difference images in 211~{\AA} and 193~{\AA} exhibit dimming in the remote but connected region, AR13602, following the flare. The CME’s position angle and velocity, measured by LASCO, together with the timing of the dimming, the EUV wave and the arrival of the CME at the base of LASCO C2 images, support the interpretation that the EUV wave is produced by the expansion of the CME in the low corona, and that the dimming in AR13602 is due to the loss of material to the expanding CME. Dynamic spectra from ORFEES and CALLISTO reveal a type II radio burst with many reverse herringbones. Some of the herringbones were also observed spatially by NRH. No forward drifting herringbones can be seen however this may be due to poorer resolution at lower frequencies.

\subsection{EUV observations}
We followed the recommended processing steps for AIA images, utilising tools from the aiapy.calibrate subpackage. For each AIA map used in our analysis, we applied the update\_pointing function to update the image metadata that describes the satellite pointing. We then applied the register function which upgrades AIA images from level 1 to level 1.5 by removing the roll angle, aligning the centre of the image with the centre of the Sun, and scaling the image to a common resolution across the channels. Lastly, we normalised each map by dividing each map by its exposure time. We also removed images with exposure times below a threshold of 1.5~s from our dataset, having noted that a relatively small number of the images had an exposure time less than 1.5~s, and these images had noticeably lower intensities than the others.

AIA captured an EUV wave, visible in 211~{\AA} running difference images between 12:09:33~UT and 12:19:33UT. The angular extent and direction of propagation can be seen in the right hand panel of Fig. \ref{fig:paths_and_clicks} and in the accompanying movie in the online version. Although the wave initially appears to travel westward, at approx.\ 12:10:45~UT the bubble appears to gain a stronger northward component. Base ratio images revealed that AR13602, to the north of the flaring active region, dims as the EUV wave begins to propagate northward.

\begin{figure*}[!htp]
    \centering
    \makebox[\linewidth][c]{%
        \includegraphics[width=1.05\textwidth]{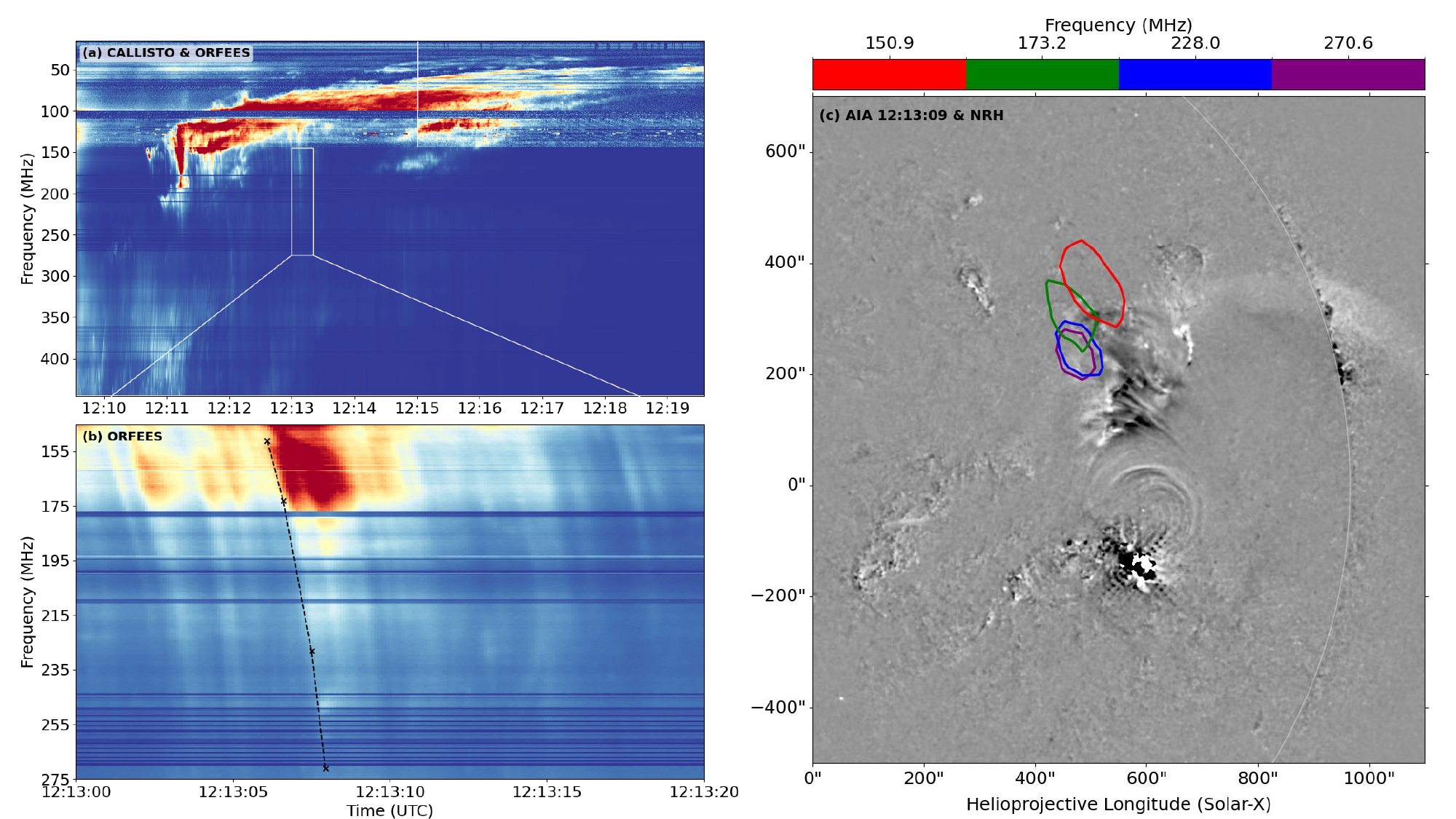}%
    }
    \caption{(a): The prolonged feature seen in the dynamic spectra is a type II radio burst captured between approx 240 and 40~MHz from approximately 12:11:30~UT to 12:20:45~UT by ORFEES and CALLISTO in Greenland and Algeria. The colour map used in this dynamic spectrum is not representative of the actual relative intensities. It was normalised differently across the two datasets with the aim of aiding the reader to best see the shape of the type II feature. (b): Zoom in on ORFEES herringbones between 12:13:00~UT and 12:13:20~UT. The points from identifying one of these herringbones are shown, at the frequencies corresponding to the NRH contours on the right (black crosses joined by a dashed black line). (c): Running difference image in 211~{\AA} from 12:13:09~UT, with NRH contours over-plotted between 150.9~MHz and 270.6~MHz. We see that the EUV wave has just passed over the open field region and reverse herringbone emission has been triggered in that region. The plotted contours show the path of the herringbone identified. In the online version we present this information in a movie.}
    \label{fig:orfees_callisto}
\end{figure*}

\subsection{Radio observations}
Between 12:11 and 12:18~UTC a type II radio burst with herringbones was observed by the CALLISTO solar spectrometer stations in Greenland and Algeria, as well as ORFEES radio spectrograph in Nançay, France (see Fig. \ref{fig:orfees_callisto}). The extended CALLISTO network, e-CALLISTO, includes stations across the world, such that at least one instrument is observing the Sun almost 24 hours a day. The total frequency range of CALLISTO is from 45 to 870~MHz and it has a time resolution of 0.25 sec. The integration time is 1 ms and the radiometric bandwidth is about 300~KHz \citep{benz}. ORFEES observes the whole-Sun flux density between 144 and 1004~MHz, which corresponds to regions between the low corona and half a solar radius above the photosphere \citep{2021JSWSC..11...57H}. Combining radio spectrogram data from ORFEES, CALLISTO Algeria and CALLISTO Greenland, we can see the type II radio burst from approximately 150 to 50~MHz and its reverse herringbones from approximately 150 to 270~MHz (see Fig.~\ref{fig:orfees_callisto}a and b). When plotting dynamic spectra we performed background normalisation to remove the background response.

The NRH captured image data during the event at 10 frequencies between 150.9 and 444~MHz, with cadence of 0.25~s. We used SolarSoftWare (SSW) to process the NRH data \citep{freeland1998data}. This involved cleaning the data with the default parameter settings and computing images. Imaging of sources at 150.9 to 270.6~MHz shows that the herringbone electron beams are localised to a specific region in the EUV wave's path, and that they are triggered by the arrival of the wave.

\begin{figure*}[!htp]
    \centering
    \makebox[\linewidth][c]{%
        \includegraphics[width=1.2\textwidth]{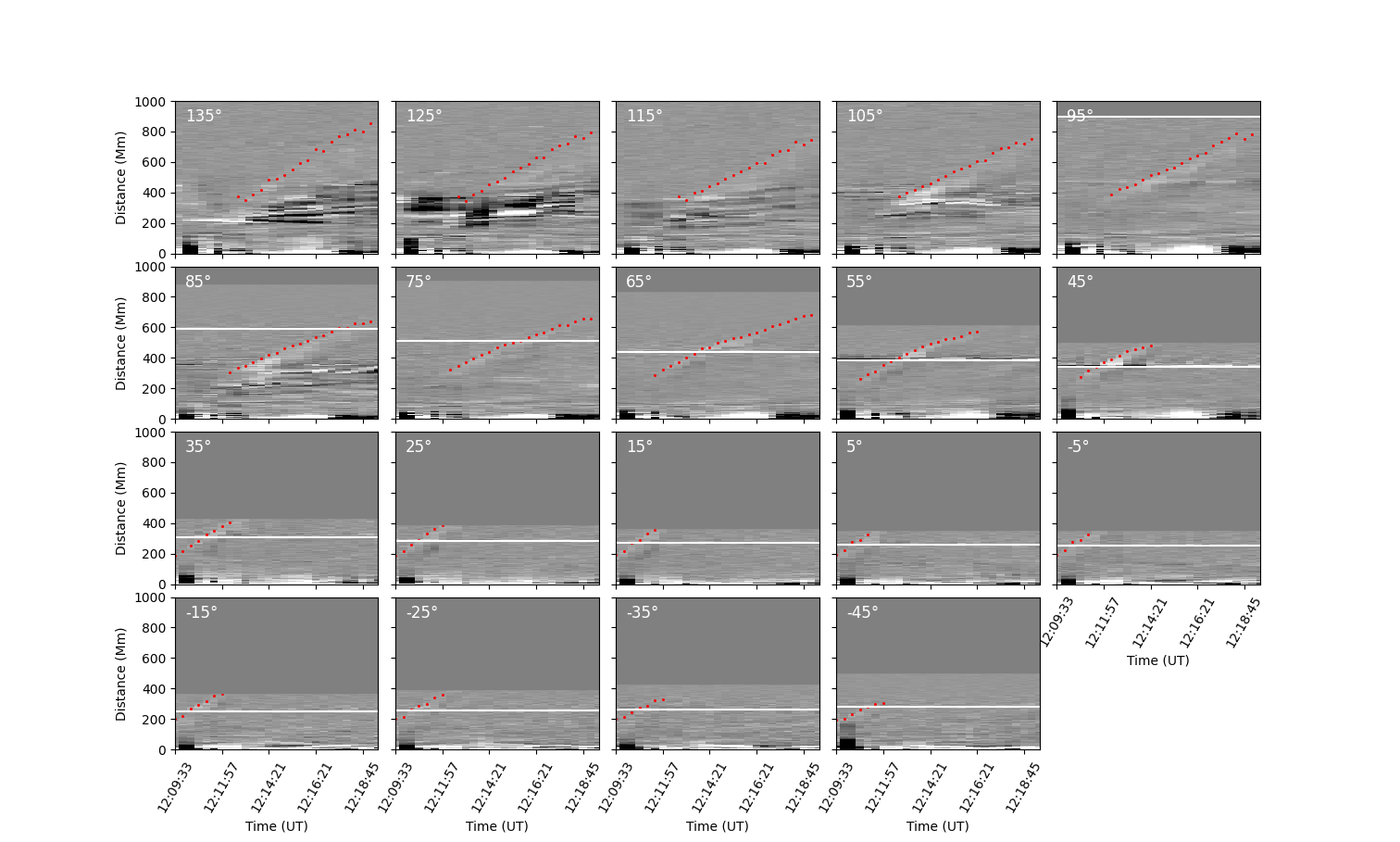}%
    }
    \caption{Stack-plots for each of the paths shown in the left hand side of Fig. \ref{fig:paths_and_clicks}. From top left to bottom right they are organised by decreasing angle. The stack-plots in the top row correspond to those along great arc paths, and the others correspond to those along straight line paths. The white horizontal line denotes the distance from AR13599 to the limb along the relevant path. For each frame where the wave is visible along the path in question, the distance from the AR of the nearest clicked point to the path is shown in red.}
    \label{fig:jplots}
\end{figure*}

\section{Methods and results}
\subsection{EUV analysis}
To characterise the global wave, we produced running difference images in 211~{\AA}, with an offset of 1 minute. From these images, we identified the angular expanse of the EUV wave to be from about -45\textdegree\ to 135\textdegree\, defining 0\textdegree as West of the NOAA Active Region 13599, at helioprojective latitude: -125", helioprojective longitude: 606". Paths starting at this point were picked at 10\textdegree\ intervals in this chosen angular range, using great arcs in the region where $\theta > 85$\textdegree and straight radial lines for all $\theta \le 85$\textdegree. This delimits the sector of the wave's path that is predominantly on disk in the field of view from that which has a significant portion off disk. On disk, we track the wave along the curved surface of the Sun, whereas off disk we track it in the plane of sky. The left panel of Fig. \ref{fig:paths_and_clicks} shows these paths plotted on the AIA 211~{\AA} running difference image for 12:11:09UT. For each path, we created a distance/time stack-plot by extracting the pixel intensities along the path for each frame over the duration of the EUV wave. Our method was similar to that used in \citet{2016SoPh..291.3217F}. These stack-plots can be seen in Fig.~\ref{fig:jplots}. The bright feature in a distance/time stack-plot has a slope that corresponds to the speed of the wave along the direction of the path. We visually identified points along the bright feature of the intensity stack-plots. We used linear fitting to find the speed of the wave along each of the chosen paths. We found a mean velocity of 1029 $\pm$ 45~\si{\kilo\metre\per\second}. We also used quadratic fits to estimate any acceleration of the wave, finding there was little to no acceleration, with a mean value of -3.6 $\pm$ 2.5~\si{\kilo\metre\per\second\squared}. For paths that passed through the dimming region, i.e.\ those labelled 115 to 135\textdegree, we found a mean velocity of 798 $\pm$ 73~\si{\kilo\metre\per\second}.

In addition to the stack-plot technique, we also employed a point-and-click method to visually identify the leading edge of the EUV wave in each running difference image. The EUV wave was visible in our 211~{\AA} running difference images between 12:09:33~UT and 12:19:33UT. For each of these images, points on the leading edge of the EUV wave were first manually identified and then joined to draw a wavefront. In the right panel of Fig. \ref{fig:paths_and_clicks}, all of these manually estimated wavefronts have been plotted on a 211~{\AA} running difference image from 12:11:09UT. In the online version, we present this in movie form. The EUV brightening was fainter at later times, when the wave was furthest from the active region source. Therefore, in these later frames it was more difficult to pinpoint the leading edge. Along the longest paths, the stack-plots also became faint far from the source. In order to compare the leading edge of the brightening that we detected in the running difference images and the stack-plots, we over-plotted on the stack-plots the clicked points closest to the paths in each frame, as seen in Fig. \ref{fig:jplots}.

We estimated the Alfv\'en Mach number, $M_A$ of the global wave as it passed through the dimming region and produced herringbones. We used the approach of  \citet{2011JASTP..73.1096Z} to estimate the density compression ratio $X$, from the observed 193~{\AA} intensity enhancement \( I/I_0 \), using the following relation:
\begin{equation}
X = \frac{n}{n_0} = \sqrt{\frac{I}{I_0}},
\end{equation}
where $I_0$ is the mean background intensity, and $I$ is the peak intensity during the passing of the wave. As is further discussed by Zhukov (2011), the assumption that $I \propto n_e^2$ is most suited to a passband whose temperature response is dominated by Fe XII emission, formed at $\mathbf{\sim}$1.4~MK, making 193{\AA} the most suitable AIA passband for this analysis.
To calculate $M_A$ from $X$ we use the equation below from \citet{2002A&A...396..673V}:

\begin{equation}
\begin{split}
(M_A - X)^2\left(5\beta X + 2M_A^2 \cos^2\theta (X - 4)\right) \\
+ M_A^2 X \sin^2\theta \left[(5 + X)M_A^2 + 2X(X - 4)\right] = 0,
\end{split}
\label{eq:MA1}
\end{equation}
where plasma $\beta$ is the ratio of the thermal plasma pressure to the magnetic field pressure, and $\theta$ is the angle between the shock normal and the magnetic field. For a parallel shock, $\theta = 0^\circ$ and Equation  \ref{eq:MA1} becomes:
\begin{equation}
M_A = \sqrt{X},
\end{equation}
and for a perpendicular shock, $\theta = 90^\circ$:
\begin{equation}
    M_A = \sqrt{\frac{X(X+5)}{2(4-X)}}.
\end{equation}
We did not directly calculate the Alfv\'en Mach number of the dimmed region as the dimming caused fluctuations in intensity which would have caused a high degree of uncertainty in the background intensity and the resultant Mach number. In order to estimate the Mach number of the wave passing through the dimmed region, we chose a region in the quiet corona that the wave passed through at approximately the same time as it passed through the dimmed region. We found the mean intensity ratio in the region, comparing its peak frame to a background frame. We calculated the resulting Alfv\'en Mach number both for the parallel and perpendicular shock configurations, finding $M_A = 1.004 \pm 0.002$ for a parallel magnetic field and $M_A =  1.006 \pm 0.003$ for a perpendicular magnetic field. These uncertainties are propagated from the standard error of the intensity ratio in the region, as we considered this to be the largest source of uncertainty.

In order to quantify the level of dimming in AR13602 at the time of the main herringbone emission in this region (12:13:00 to 12:13:20~UT), we produced base difference ratio images from AIA 211~{\AA} images, using a pre-flare base image from 12:08:45UT. The pixel intensities of a base difference ratio image, $I_{bdr}$, are
\begin{equation}
    I_{bdr} = \frac{I - I_0}{I_0}
\label{bdr}
\end{equation}
where $I_0$ is the pixel intensity in the base image and $I$ is the pixel intensity in the current image. We examined the time profile of mean pixel values of base difference ratio images, and found that at the time of interest, the mean intensity of the dimming region had dropped by approximately $45\%$ of its base level intensity. For more detail see Fig. \ref{fig:dimming}. As detailed in \citep{2010ApJ...720L..88R}, dimming in coronal pass bands can occur due to a decrease in plasma density or a change in plasma temperature. From base difference images, we identified simultaneous dimming in AR13602 in the 211~{\AA}, 193~{\AA}  and 171~{\AA}  passbands and from running difference images we did not identify any enhancement in 94~{\AA}. This combination of multi-channel dimming and the absence of signatures of heating disfavours a thermal interpretation and favours density depletion due to plasma evacuation. Scaled Newkirk models with fold numbers between 2 and 4 are routinely used to approximate the radial density profile of active regions \citep[e.g.][]{2014SoPh..289..251V, 2018A&A...611A..57M, 2025A&A...695A.136B}. Attributing the dimming of this active region by 45\% to a decrease in density, we applied a Newkirk density model with scaling factors of 1.3-2.6, to approximate the region's density profile.

\begin{figure}
    \includegraphics[width = 0.5\textwidth]{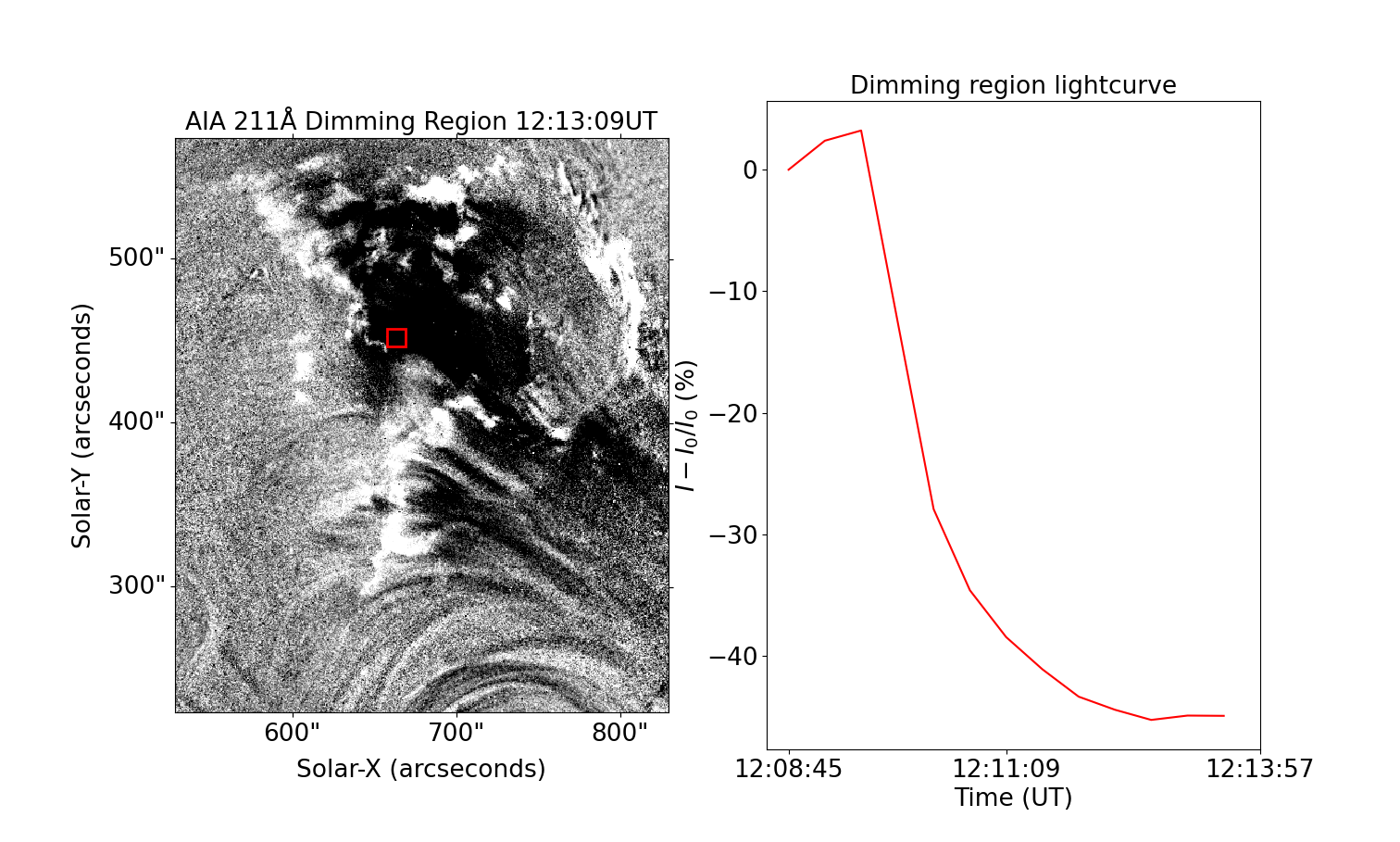}
      \caption{Left: 211 {\AA} base difference ratio image of the area surrounding AR13602, at 12:13:09UT, using a quiet time base frame from 12:08:45UT. The red box indicates the subregion that the light-curve on the right corresponds to. This subregion was identified as consistently being within the dimming region during the time window that dimming occurred. Right: The base difference ratio light curve for the subregion, showing that it dims by up to 45.77\%.}
    \label{fig:dimming}
\end{figure}

 \begin{figure*}
\sidecaption
  \includegraphics[width = 0.6\textwidth, trim=0 0 200 0, clip=]{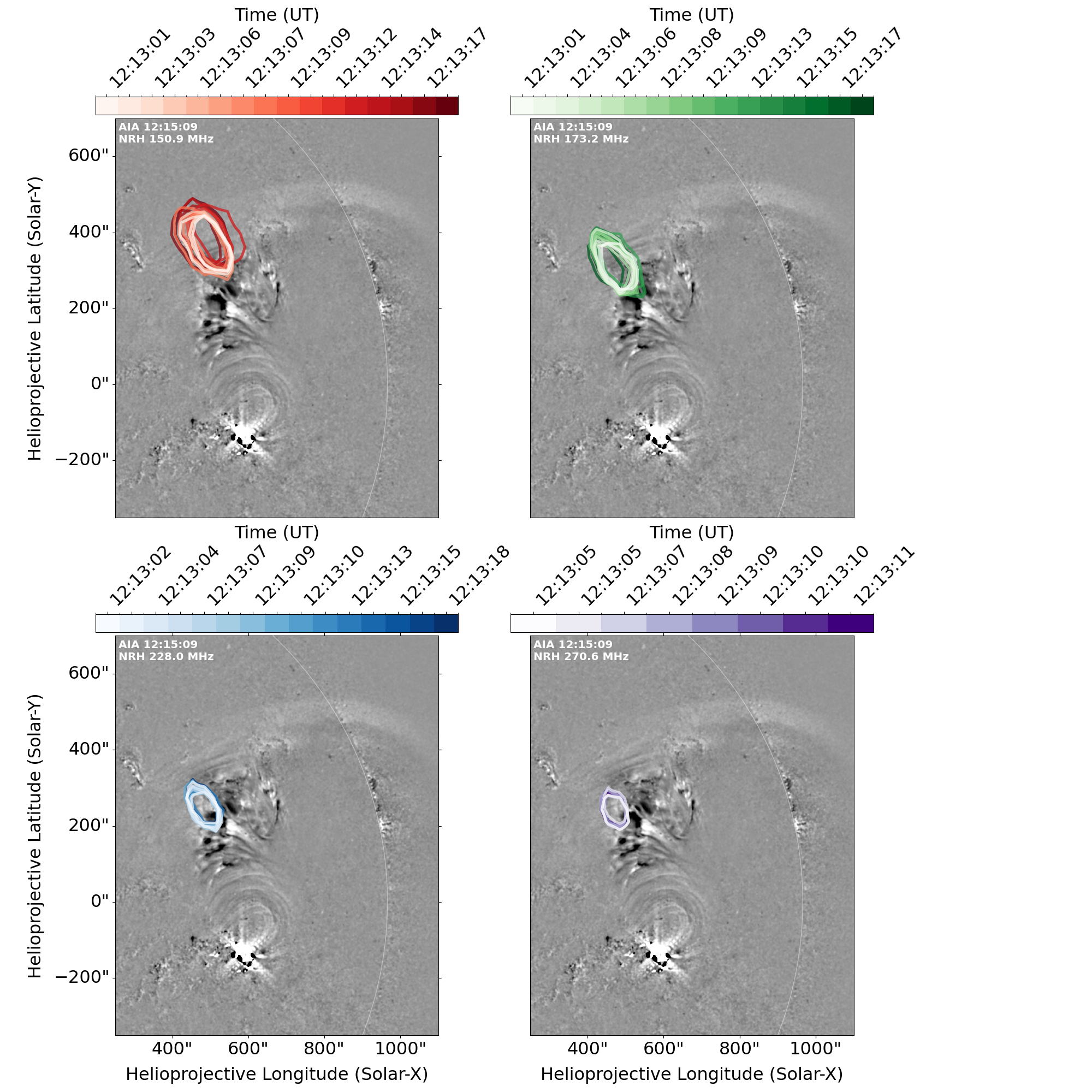}
     \caption{From top left to bottom right the NRH contours that correspond to herringbone emission between 12:13:00~UT and 12:13:20~UT at 150.9~MHz, 173.2~MHz, 228.0~MHz and 270.6~MHz, over-plotted on the 211~{\AA} running difference image for 12:15:09UTC. For 150.9~MHz and 173.2~MHz there are sixteen contours corresponding to the sixteen herringbones. Fifteen herringbones extend to 228.0~MHz, so fifteen contours are shown in the bottom left panel. Only eight herringbones extended down to 270.6~MHz, hence eight contours are shown on the bottom right panel.}
     \label{fig:hb_nrh}
\end{figure*}

We also estimated the Alfv\`en Mach number with a second method for comparison. We first estimated the Alfv\`en speed in the dimming region based on typical active region values, and then calculated the Mach number as the ratio of the speed of the wave in this area (798~\si{\kilo\metre\per\second}) to the Alfv\`en speed. Assuming a height of 1.1~\(R_\odot\) (approximately the typical height of coronal waves \citep[cf.][]{STEREO-euvwave-height}), and adopting a coronal magnetic field strength of 10~G, consistent with empirical active region coronal field models presented in \citep{1999SoPh..186..123G}, together with a 1.3 fold Newkirk density model, we calculate an Alfv\`en speed of 900~\si[detect-weight=true,detect-family=true]{\kilo\metre\per\second}, thus a Mach number of 0.89. If we use a 2.6 fold Newkirk model, we find an Alfv\`en speed of 637~\si[detect-weight=true,detect-family=true]{\kilo\metre\per\second}, thus an Alfv\`en Mach number of 1.25.

\subsection{Radio analysis}
For each NRH frequency we over-plotted the 80 and 90\% contours on the AIA 211~{\AA} running difference images and collated these in a movie, with the dynamic spectra from ORFEES and CALLISTO for the same time window displayed along side it. From the movie we noticed the following. A group of reverse drifting herringbones from 12:11:12~UT to 12:11:20~UT appears to be triggered at the southern edge of the dimming region, AR13602, predominantly between 150.9 and 228.0~MHz. This coincides with the arrival of the EUV wave in this region. At 12:12:27~UT the dimmed region again becomes the main source of 150.9 - 173.2~MHz radio emission. Between 12:12:27~UT and 12:14:13~UT, this source moves along with the EUV wave, at times extending to 228.0~MHz. This implies that the emission is due to accelerated particles at the shock front. At 12:13:05~UT the 270.6~MHz contours become aligned with this source. 

We identified sixteen reverse herringbones in the ORFEES dynamic spectrum between 12:13:00~UT and 12:13:20~UT, drifting from approx. 150~MHz to approx. 270~MHz. NRH images reveal that this emission is localised to the dimmed region. Fig. \ref{fig:orfees_callisto} highlights one herringbone in the dynamic spectra and shows its closest NRH contours over-plotted on an AIA 211~{\AA} running difference image. For each of the sixteen herringbones, we identified the frequency/time points on each herringbone that matched the four NRH imaging channels: 150.9, 173.2, 228.0 and 270.6~MHz. Fig. \ref{fig:hb_nrh} shows the position in time of the herringbone electron beams, at the heights corresponding to the four channels.

Herringbones have been approximated by straight lines in previous works, \citet[e.g.][]{2015AGUFMSH22B..01C, 2024A&A...683A.123Z}. The herringbones in this event had steepening drift rates that were better approximated by quadratic curves than straight lines. Using Equation \ref{speed2}, and scaled Newkirk density models to relate $f_p$ to a corresponding radius, these drift rates imply radial beam speeds that increase as the electrons travel deeper into the corona. This is due to the fact that the electrons are travelling downwards along curved magnetic field lines. The speed of these particles is not in fact increasing as they travel downwards, but the radial component of their velocity is increasing as the field lines they travel on become more purely radial lower down.

Given the relationship between plasma density ($n_e)$ and plasma frequency ($f_p)$, $f_p = C\sqrt{n_e}$, coronal electron density models allow us to estimate the height of radio emission based on its frequency, assuming that it is either fundamental emission ($f = f_p$) or harmonic emission ($f = 2f_p$). The frequency drift of herringbones indicates the propagation of the beam through areas of changing density. Assuming that this change in density corresponds to a change in coronal height in accordance with a density model, $n_e(r)$, we can infer the speed of the beam from the drift rate ($df/dt$) using the following expression \citep{2019NatAs...3..452M},
\begin{equation}
v = \frac{2 \sqrt{n_e}}{C} \left( \frac{dn_e}{dr} \right)^{-1} \frac{df}{dt}.
\label{speed}
\end{equation}
Using a \citet{1961ApJ...133..983N} density model of the form:
\begin{equation}
n_e(r) = N_0\cdot10^{\alpha/r},
\label{model}
\end{equation}
Equation \ref{speed} becomes 
\begin{equation}
v = \frac{-2r^{2}}{\alpha\ln{10} f_p} \frac{df}{dt},
\label{speed2}
\end{equation}
where $r$ is chosen to be the starting height of the herringbone and $f_p$ is chosen to be the starting frequency.

For each of the sixteen herringbones, we identified twenty points and fitted a quadratic curve to those points. Differentiating frequency as a quadratic function of time, we found drift rate as a linear function of time for each burst, and then used Equation \ref{speed2} to estimate the electron beam velocity as a function of time. We found that the herringbones had starting frequencies of approximately 150~MHz and drift rates ranging from $+$45~\si{\mega\hertz\per\second}to $+$85~\si{\mega\hertz\per\second}. The \citet{1961ApJ...133..983N} model (i.e. Equation \ref{model} with $N_0 = 4.2\times 10^{4}$, $\alpha$ = 4.32) approximates the electron density in the corona from 1-3~\(R_\odot\), with different fold numbers depending on the activity of the region. We found upper and lower limits for the herringbone electron speeds and starting heights, based on scaling factors of 1.3 to 2.6. Given this range of scaling factors, if we assume that the radio emission is fundamental, the electron beams have a starting height in the range of 1.16 to 1.27~\(R_\odot\), and have mean velocities of 0.18 to 0.21 $c$ , corresponding to mean kinetic energies of 8.02 to 11.67~keV. If the emission is assumed to be harmonic, using the same range of scaling factors, we get a starting height in the range of 1.39 to 1.54~\(R_\odot\), downward velocities of 0.49 to 0.59 $c$, and kinetic energies of 75.32 to 122.10~keV.

\subsection{Magnetic field}
We modelled the pre-eruption coronal magnetic field by performing a potential field source surface extrapolation with the sunkit-magex.pfss Sunpy module \citep{Stansby2020}, using the Global Oscillation Network Group (GONG) magnetogram synoptic map observations from 12:04:00~UT. Fig. \ref{fig:pfss} shows the PFSS field lines overplotted on the closest AIA 193{~\AA} image at 12:04:04~UT. For the PFSS extrapolation we made a grid of 20 by 20 seed footpoints in the range of -85° to 85° in latitude and 0° to 85° in longitude, at a height of 1.15~\(R_\odot\). The apparent extension of field lines beyond this region in Fig. 5 arises because field lines traced from these seeds may map to different locations outside these bounds. We chose these ranges as the EUV wave is only visible propagating across the Western half of the disk, and we wanted to avoid the polar regions, where uncertainties in the extrapolated field become very large.

\begin{figure}[!t]
      \includegraphics[width =  0.5\textwidth]{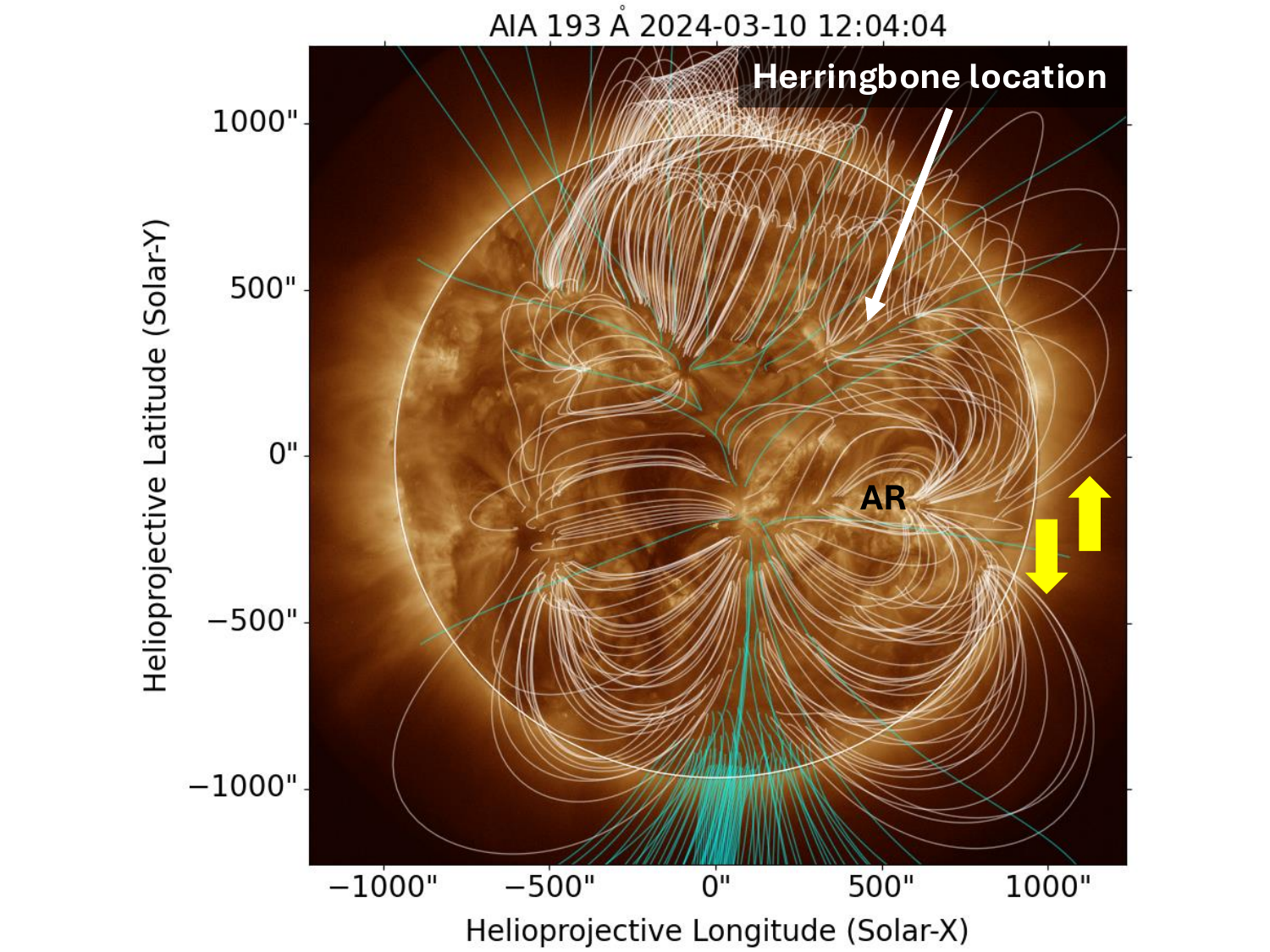}
      \caption{The Potential Field Source Surface (PFSS) magnetic field extrapolation just before the flare. The field lines are plotted on an AIA 193~{\AA} image. The closed field lines are in white and open field lines are in cyan. The location of the herringbone contours in AR13602 is as indicated. The yellow arrows highlight the large loop systems to the North and South of the flaring active region, which is marked by the `AR' label.}
    \label{fig:pfss}
\end{figure}
Fig. \ref{fig:pfss} shows that there was a large loop system to the East and South of AR13599, and another to the North of the active region, connecting AR13599 to AR13602. These loops indicate regions of strong, closed magnetic fields in the low corona. Such regions have higher Alfv\`en speeds, reducing the Alfv\`en Mach number of the wave and diminishing its ability to compress and heat plasma. This caused the bright front to initially propagate preferentially towards the West. The gap between the 2 loop systems is approximately the expanse of the EUV wave in the first few frames of our 211{~\AA} running difference movie, as seen in the right panel of Fig. \ref{fig:paths_and_clicks}. Coronal dimming in AR13602, apparent in 211{~\AA} base difference images, suggests that material flows out from the region into the heliosphere, leaving behind an area of open magnetic field. Due to this change in the configuration of the magnetic field, the EUV wave then propagates in a more Northerly direction.

\section{Discussion and conclusions}
In this paper, we have presented observational evidence of localised particle acceleration by a weak, subcritical coronal shock associated with the 10 March 2024 eruption. This event provides a useful set of observations to study the conditions in which a weak plasma shock can give rise to particle acceleration. 

By analysing AIA images, and modelling the pre-eruption coronal magnetic field, we examined the propagation path of the wave and the coronal conditions it encountered. We found that during the passage of the wave, there was an outflow of plasma from a connected active region, which was related to a reconfiguration of the magnetic field. This allowed the EUV wave, which initially travelled approximately due West, to gain a more Northward component. The EUV wave then propagated into the region from which material had been lost, where it encountered open magnetic field lines, quasi-perpendicular to its path. A full understanding of the trajectory of the EUV wave and the evolution of the erupting flux rope, which was not directly observable, would require dedicated magnetohydrodynamic modelling of the eruption and its interaction with the surrounding coronal magnetic field.

The type II radio burst and reverse drifting herringbones seen in the dynamic spectra indicate acceleration of particles by a shock. Radio images from NRH confirm that this particle acceleration occurs at the open field region, when the EUV wave impacts this open field. We checked the location of each radio contour for sixteen herringbones, at the four relevant frequency channels of NRH, finding that these electron beams were undoubtedly localised in the North East of the dimming region, and that they were temporally confined to the window when the EUV wave passed through this region. (Fig.\ref{fig:hb_nrh}).

The combined radio and EUV observations indicate that the wave encounters conditions favourable for shock particle acceleration when it reaches the remaining open field lines in the dimmed region. From the increase in 193~{\AA} intensity at the EUV wavefront, we estimate an Alfv\`en Mach number of $M_A \approx 1.004 - 1.006$ depending on the angle between the ambient magnetic field lines and the shock normal. These values are within the range of Alfv\`en Mach numbers that we calculate based on the speed of the EUV wave through the dimmed region (798~\si{\kilo\metre\per\second} ) and typical Alfv\`en speeds, considering that the dimming region is also an active region. Assuming a height of 1.1$R_{\odot}$ and magnetic field strength of 10~G \citep{1999SoPh..186..123G}, and approximating the electron density using a 1.3 – 2.6 $\times$ Newkirk density model, we found a broader range of $M_A \approx 0.89 - 1.25$. This is a rough estimate due to our assumptions, however it is consistent with the values we calculated from the intensity enhancement, both methods leading us to the conclusion that the EUV wave was weakly shocking when it passed through AR13602.

Despite this shock being subcritical, the fast drifting herringbones seen in the dynamic spectra imply that electrons were accelerated at the shock front to weak relativistic energies. Using a 1.3-2.6 fold Newkirk density model and the assumption of fundamental emission, we find herringbone electron velocities between 0.18 and 0.21 $c$, corresponding to kinetic energies between 8.02 and 11.67~keV. These values are within the range found in previous studies such as \citet{2005A&A...441..319M} and \citet{2013NatPh...9..811C}. If the emission is assumed to be harmonic we find velocities of 0.49 to 0.59 c, and kinetic energies of 75.32 to 122.10~keV, which is outside of the normal range found in previous studies. We found from the ACE/EPAM detections that electron flux in the 38-53keV, 53-103keV, 103-175keV and 175-315keV ranges were enhanced within the hour after the herringbones were observed. This is consistent with the expected time of arrival of the herringbone electrons at L1 and is also consistent with the calculated electron energies if the herringbone emission is assumed to be harmonic. From the WIND/3DP detections of electrons with energies $<$25keV, an enhancement was not observed in the time window or in the energy range that we would expect if we assume the herringbone emission to be fundamental. These observations favour the interpretation of the radio emission as harmonic.

The appearance of localised reverse herringbones as the EUV wave encountered the region of open field suggests that the local magnetic field geometry played a key role in the acceleration of electrons. The presence of a shock, albeit a very weak one, combined with perpendicular open field lines, allowed electrons to be efficiently energised here via shock drift acceleration, and to escape the shock region along field lines, producing the observed herringbones.

Our findings support the theory that shock drift acceleration is particularly efficient when the shock normal is quasi-perpendicular to the ambient magnetic field \citep{2013NatPh...9..811C}. Localised radio emission from a weak coronal shock interacting with magnetic structures has been previously reported by \citet{2021ApJ...921...61L}. Our results add to mounting evidence that subcritical shocks can accelerate electrons to keV energies if the magnetic geometry is favourable. This implies that the efficiency of particle acceleration in the low corona is strongly affected by the local magnetic field configuration.
Future coordinated EUV and radio imaging will be key to quantifying how frequently such conditions occur and to assessing their contribution to space weather.

\begin{acknowledgements}
      We would like to thank Diana Morosan, Astrid Veronig, Jasmina Magdalenic and Laura Hayes for their contributions to this work in the form of helpful discussions regarding our analysis. We also thank the anonymous referee for their constructive comments which helped to improve the paper. We acknowledge the use of data from the SDO/AIA instrument. We thank the Radio Solar Database service at LESIA (Observatoire de Paris) for providing the NRH and ORFEES data, and the e-Callisto network and ETH Zurich, for providing Callisto spectrometer data. This research used version 6.0.2\footnote{https://doi.org/10.5281/zenodo.13743565} of the \textit{SunPy} open source software package \citep{sunpy_community2020}, and version 0.7.4 of the aiapy open source software package \citep{2020JOSS....5.2801B}. This work also made use of Astropy:\footnote{http://www.astropy.org} a community-developed core Python package and an ecosystem of tools and resources for astronomy \citep{astropy:2013, astropy:2018, astropy:2022}. This work is supported by the project "The Origin and Evolution of Solar Energetic Particles”, funded by the European Office of Aerospace Research and Development under award No. FA8655-24-1-7392.

\end{acknowledgements}

\bibliographystyle{aa}
\bibliography{refs}

\end{document}